\documentclass[aps,prb,reprint,showpacs,amsmath,amssymb]{revtex4-1}
\usepackage{graphicx}
\usepackage{bm}

\begin{document}

\title{Spin relaxation in a zinc-blende (110) symmetric quantum well with delta-doping} 

\author{Hiroshi Akera, Hidekatsu Suzuura, and Yoshiyuki Egami}

\affiliation{Division of Applied Physics, Faculty of Engineering, 
Hokkaido University, Sapporo, Hokkaido, 060-8628, Japan}

\date{\today}

\begin{abstract}
The spin relaxation of a two-dimensional electron system (2DES)
formed in a symmetric quantum well 
is studied theoretically 
when the quantum well is parallel to the (110) plane of the zinc-blende structure,  
the spin polarization is perpendicular to the well, 
and electrons occupy only the ground subband.  
The spin relaxation rate is calculated as a function of 
the distribution of donor impurities which are placed in the well layer. 
Considered processes of the spin relaxation are 
(1) intrasubband process 
by impurity-potential-induced spin-orbit interaction (SOI), 
which is the Elliott-Yafet mechanism in the 2DES, 
and 
(2) virtual intersubband processes consisting of 
a spin flip by 
(2a) well-potential-induced SOI 
or 
(2b) the Dresselhaus SOI,  
and a scattering from an impurity. 
It is shown that all of the above processes disappear, 
when all impurities are located on the center plane of the well. 
Even if impurities are distributed over three (110) atomic layers, 
the spin relaxation rate is two orders of magnitude lower than   
that for the uniform distribution over the well width of 7.5\ nm. 
In GaAs/AlGaAs type-I quantum wells 
the processes (1) and (2a) interfere constructively,  
being dominant over (2b) for the well width of $\sim$10\ nm, 
while in some type-II quantum wells 
they can interfere destructively. 
\end{abstract}



\maketitle

\def\ve{\varepsilon}
\def\Te{T_{\rm e}}
\def\muec{\mu_{\rm ec}}
\def\kB{k_{\rm B}}
\def\Vec#1{\bm{#1}}

\section{Introduction}

Employing the spin degree of freedom in semiconductors 
is a promising approach to the development of hybrid devices which perform 
all of information processing, communications, and storage.\cite{Awschalom2007}
The prerequisite spin polarization can be created in nonmagnetic semiconductors 
by the spin-orbit interaction (SOI) 
through the spin Hall effects of the extrinsic origin\cite{Dyakonov1971JETPL, Dyakonov1971PhysicsLettersA, Hirsch1999, Zhang2000, Engel2005} 
and of the intrinsic one,\cite{Murakami2003, Sinova2004} 
which have been confirmed in experiments.\cite{Kato2004, Wunderlich2005, Sih2005} 
However, the same SOI becomes 
a driving force of the spin relaxation in various mechanisms.\cite{Zutic2004spintronics,Wu2010} 
In this paper we show theoretically that 
the spin relaxation due to a spin flip by the SOI with a scattering at an impurity
vanishes for a two-dimensional electron system (2DES) 
formed in a symmetric quantum well with a delta-doping ($\delta$-doping)\cite{Ploog1987,Haug1987} 
on the center plane of the well. 

Two major mechanisms of the spin relaxation in $n$-doped semiconductors are 
the Dyakonov-Perel mechanism\cite{Dyakonov-Perel1971,Dyakonov-Perel1972,Dyakonov1986} 
and the Elliott-Yafet mechanism.\cite{Elliott1954, Yafet1963, Chazalviel1975} 
The Dyakonov-Perel mechanism is due to 
the spin precession around a SOI-induced effective magnetic field 
whose direction and magnitude depend on the momentum of each electron. 
In addition to the Dresselhaus SOI\cite{Dresselhaus1955} 
due to the inversion asymmetry in the crystal structure, 
the Rashba SOI\cite{Rashba1960,Ohkawa1974,Bychkov1984a,Bychkov1984b} 
produces the effective magnetic field 
in a 2DES formed in a quantum well with the inversion asymmetry,   
and the Dyakonov-Perel mechanism due to such SOIs 
is a major mechanism of the spin relaxation. 
Fortunately the Dyakonov-Perel mechanism can be turned off 
for the spin direction perpendicular to the 2DES  
by preparing a symmetric quantum well on a substrate 
oriented parallel to the (110) plane of the zinc-blende structure.  
This is because 
the Dresselhaus SOI in symmetric quantum wells parallel to the (110) plane 
gives an effective magnetic field perpendicular to the 2DES
regardless of electron momentum\cite{Dyakonov1986,Winkler2004} 
and the Rashba SOI is absent in symmetric quantum wells.  
 
The suppression of the spin relaxation in (110) symmetric quantum wells 
has been observed for the first time by Ohno {\it et al.}\cite{Ohno1999} 
in the pump-probe method: 
the spin relaxation time in GaAs (110) symmetric quantum wells 
is more than an order of magnitude longer than that in (100) quantum wells.
\footnote{
The reduction of the spin relaxation 
in (110) quantum wells compared to (100) quantum wells 
has also been observed\cite{Hall2003} in InAs/GaSb system, a typical type-II superlattice 
with use of the pump-probe technique.
} 
The spin relaxation remaining in their undoped sample 
was ascribed to the Bir-Aronov-Pikus mechanism\cite{Bir1975} 
due to the electron-hole exchange interaction. 
Holes are introduced in the pump-probe experiment when the sample is excited optically 
for the purpose of generation and detection of the spin polarization.  
The Bir-Aronov-Pikus mechanism, however, can be neglected 
in the later measurement by M\"uller {\it et al.}\cite{Muller2008} 
with use of the spin noise spectroscopy 
which can avoid the introduction of holes. 
Since the quantum well used in this spin noise measurement was modulation-doped, 
the observed low spin-relaxation rate of (24ns)$^{-1}$ was attributed to 
the Dyakonov-Perel mechanism due to the random Rashba field 
produced by density fluctuations of donors located in barrier layers.\cite{Sherman2003} 
In this paper we instead consider a doping in a well layer, 
and therefore the random Rashba field is outside the scope of this paper. 

A doping in a well layer 
has the advantage of efficient generation of the spin polarization 
by the extrinsic spin Hall effect. 
In fact, the spin accumulation produced by the spin Hall effect  
has been observed in AlGaAs (110) quantum wells,  
in which Si donors are doped uniformly in the well layer.\cite{Sih2005} 
The observed spin Hall effect has been explained by 
the theory of the extrinsic spin Hall effect,\cite{Hankiewicz2006PRB, Tse-Das_Sarma2006}
in which donor impurities in the well layer play a major role 
in creating the spin polarization. 

Such previous studies suggest that one promising way to achieve large spin polarizations 
is to employ an $n$-doped (110) symmetric quantum well, 
which can produce the spin polarization by the extrinsic spin Hall effect 
and, at the same time, can avoid the spin relaxation due to Dyakonov-Perel mechanism. 
An important task in this direction will be 
to find a method to suppress the spin relaxation caused 
by donor impurities introduced in the well layer. 

It is known that impurities give rise to 
the spin relaxation called the Elliott-Yafet mechanism, 
in which spin-flip scatterings are caused by 
the combined action of the impurity potential and the SOI. 
This mechanism is likely to be dominant 
for the relaxation of the spin polarization perpendicular to the 2DES 
in a (110) symmetric quantum well, 
in which the Dyakonov-Perel mechanism does not work. 
In quantum wells, the Elliott-Yafet mechanism is modified 
by the subband structure:   
in addition to intrasubband spin-flip processes,\cite{Averkiev2002, Bronold2004}
intersubband spin-flip processes
due to SOI matrix elements between states in different subbands 
contribute to the spin relaxation. 

The importance of such intersubband processes in various spin dynamics 
has been suggested 
in recent studies. 
D\"ohrmann {\it et al.}\cite{Dohrmann2004} have proposed 
a spin-relaxation mechanism due to 
intersubband spin-flip transitions, 
which are induced by the Dresselhaus SOI and the impurity potential, 
between the ground subband and the first-excited subband 
to explain their observed result of 
the spin relaxation time in a (110) symmetric quantum well 
at higher temperatures such that the first-excited subband is occupied by electrons. 
Bernardes {\it et al.}\cite{Bernardes2007} have studied theoretically 
roles of the intersubband matrix element of the SOI induced by the well potential 
in a symmetric quantum well and 
have derived the formula of the spin Hall conductivity in this system. 
Zhou and Wu\cite{Zhou2009} have calculated the spin relaxation time 
of the 2DES occupying only the ground subband in a (110) symmetric quantum well
by considering a virtual intersubband process through the first-excited subband 
in terms of the Dresselhaus SOI with the impurity potential. 

In this paper we study theoretically the spin relaxation  
in an $n$-doped (110) symmetric quantum well 
for the spin orientation perpendicular to the well. 
We consider the 2DES occupying only the ground subband 
and study spin-flip scatterings 
through both intrasubband and intersubband processes. 
The intrasubband spin-flip scattering is caused by 
the SOI due to the impurity potential. 
The intersubband spin-flip scattering is 
a virtual process through one of excited subbands, 
which consists of 
an intersubband spin-flip process due to the SOI 
and an intersubband scattering process due to the impurity potential.\cite{Zhou2009} 
We take into account both the well-potential induced SOI and the Dresselhaus SOI 
for the intersubband spin-flip process.  
In particular, we investigate the dependence of the spin-flip scattering rate 
on the position of delta-doping,\cite{Ploog1987,Haug1987} 
which can introduce impurities within an atomic layer in the well. 

The organization of the paper is as follows. 
In Sec.~\ref{sec:Hamiltonian}, we describe the Hamiltonian, 
which includes the SOIs originating from the impurity potential and the well potential  
in addition to the Dresselhaus SOI. 
In Sec.~\ref{sec:center_delta-doping}, we show that spin-flip scatterings are absent 
when impurities are placed only in the atomic layer at the center of the well 
(center delta-doping). 
In Sec.~\ref{sec:off-center_delta-doping}, 
we investigate spin-flip scatterings for off-center delta-dopings 
by calculating the spin-flip scattering rate 
as a function of the position of delta-doping.  
We also calculate the spin-flip scattering rate 
for impurity distributions having nonzero widths. 
In Sec.~\ref{sec:conclusions}, conclusions are given.
 
\section{Hamiltonian}
\label{sec:Hamiltonian}

\begin{figure}[ht]
\vskip -0.8cm
\includegraphics[width=9.5cm, bb=0 0 595 842]{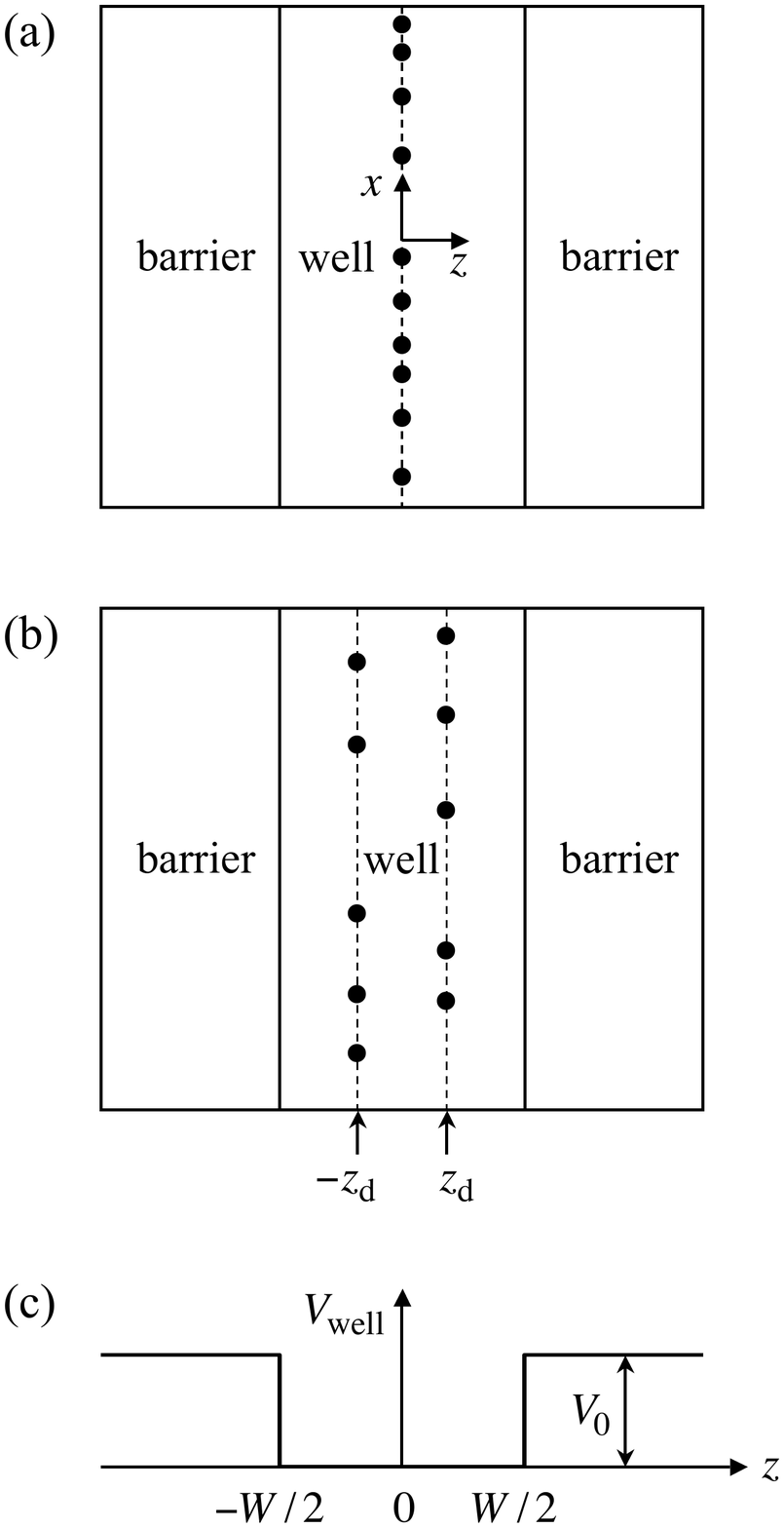}
\vskip -1.3cm
\caption{\label{fig:delta_doping}
(a) Delta-doping at the well center ($z=0$). Filled circles ($\bullet$) represent impurities. 
(b) Delta-dopings at $z=\pm z_{\rm d}$. 
(c) The well potential, $V_{\rm well}(z)$, with height $V_0$ and width $W$. 
}
\end{figure}

We consider electron states in a quantum-well structure 
which is formed by two different zinc-blende semiconductors. 
The Hamiltonian is 
\begin{equation}
H = H_0 + H_1 .
\end{equation}
The unperturbed Hamiltonian $H_0$ is
\begin{equation}
H_0= \frac{\Vec {\hat p}^2}{2m} + V_{\rm well}(z) ,
\end{equation}
where $\Vec {\hat p}=(\hat p_x, \hat p_y, \hat p_z)=
-i \hbar \Vec \nabla=-i \hbar(\nabla_x, \nabla_y, \nabla_z)$ and 
$m$ is the effective mass of the conduction band.   
The well potential $V_{\rm well}(z)$, 
which is the potential due to the offset of the conduction band 
at the interface between two constituent semiconductors,
\footnote{
The boundary conditions for the conduction-band envelope function $\varphi(z)$
in the $\Vec k \cdot \Vec p$ theory 
 for heterostructures,\cite{Lassnig1985, Bernardes2007,Calsaverini2008} 
which employs a single Hamiltonian with $z$-dependent band edges, 
are the continuity of $\varphi(z)$ and that of $m(z)^{-1}\nabla_z \varphi(z)$ at each interface 
where $m(z)$ is the effective mass of the conduction band. 
This so-called envelope-function approximation is shown\cite{Ando1989} 
to be accurate in GaAs/AlGaAs heterostructures. 
In this paper we neglect the difference in the effective mass between GaAs and AlGaAs 
following Ref.\ \onlinecite{Bernardes2007}. 
}  
is given for the width $W$ and the height $V_0(>0)$ by 
\begin{equation}
V_{\rm well}(z)=
\left\{ \begin{array}{ll}
                           0      &(|z|<W/2)  ,  \\ 
                           V_0  &(|z|>W/2)  , 
           \end{array}   \right. 
\end{equation}
and is illustrated in Fig.\ \ref{fig:delta_doping}. 
Each eigenstate of $H_0$ is labelled by 
the subband index, $n=0,1,2,\cdots$, 
and the wave vector in the $xy$ plane, $\Vec k=(k_x,k_y)$, 
and the $z$ component of spin, $\sigma=\uparrow, \downarrow$. 
The corresponding eigenenergy depends only on $n$ and $k=|\Vec k|$ 
and is denoted by $\ve_{nk}$ or $\ve_{n \Vec k}$.  
We assume that only the ground subband with $n=0$ is occupied by electrons.   

The perturbation $H_1$ is 
\begin{equation}
H_1= V_{\rm imp}(\Vec r) 
+ H^{\rm so}_{\rm imp} + H^{\rm so}_{\rm well} + H^{\rm so}_{\rm D} . 
\end{equation}
Here $V_{\rm imp}(\Vec r)$ with $\Vec r=(x,y,z)$
 is the potential due to randomly-distributed impurities.  
$H^{\rm so}_{\rm imp}$ is the SOI due to $V_{\rm imp}(\Vec r)$, 
given by 
\begin{equation}
H^{\rm so}_{\rm imp}= - \frac{\eta}{\hbar} {\Vec \sigma} \cdot 
\left(\Vec \nabla V_{\rm imp} \times {\Vec{\hat p}} \right) ,
\label{eq:SOI_imp}
\end{equation}
while $H^{\rm so}_{\rm well}$ is that caused by 
the well potential for an electron in each of the valence bands,  
defined by 
\begin{equation}
H^{\rm so}_{\rm well}= - \frac{\eta b_{\rm off}}{\hbar} {\Vec \sigma} \cdot 
\left(\Vec \nabla V_{\rm well} \times {\Vec{\hat p}} \right) ,
\label{eq:SOI_well}
\end{equation}
where ${\Vec \sigma}=(\sigma_x,  \sigma_y, \sigma_z)$ is the Pauli spin matrix and  
$\eta$ is the effective coupling constant of the SOI   
for an electron in the conduction band of the semiconductor in the well layer.  
The factor $b_{\rm off}$ is a dimensionless constant reflecting  
the difference in the band offset 
between the conduction band and each of the valence bands. 
The formula of $b_{\rm off}$ is given in Appendix. 
The last term of $H_1$ is the Dresselhaus SOI in the zinc-blende structure 
\begin{equation}
H^{\rm so}_{\rm D}= - \frac{\gamma}{2\hbar^3} {\Vec \sigma} \cdot {\Vec h}({\Vec{\hat p}}) ,
\end{equation}
where $\gamma$ is the coupling constant of the Dresselhaus SOI 
and ${\Vec h}=(h_x,h_y,h_z)$ can be understood as an effective magnetic field.  
In (110) quantum wells ${\Vec h}$ is given by 
\begin{equation}
\begin{split}
h_x&=(-\hat p_x^2 -2\hat p_y^2 +\hat p_z^2) \hat p_z , \\
h_y&=4 \hat p_x \hat p_y \hat p_z , \\
h_z&= \hat p_x (\hat p_x^2 -2\hat p_y^2 -\hat p_z^2) ,
\end{split}
\end{equation}
where the Cartesian unit vectors are taken as
\begin{equation}
\begin{split}
\Vec e_x&= ( - \Vec e_{[100]} + \Vec e_{[010]} ) / \sqrt 2 , \\
\Vec e_y&= \ \Vec e_{[001]} \ , \\
\Vec e_z&= (  \Vec e_{[100]} + \Vec e_{[010]} )  / \sqrt 2 
,
\end{split}
\end{equation}
with $\Vec e_{[100]}$, $\Vec e_{[010]}$ and $\Vec e_{[001]}$  
the unit vectors along the crystal axes.

\section{Absence of spin-flip scatterings \\ in the center delta-doping}
\label{sec:center_delta-doping}

Typical impurity-doping profiles are 
the uniform doping and the modulation doping. 
The uniform doping in a well layer has been used  
in measurements\cite{Ohno1999, Sih2005} and a calculation\cite{Averkiev2002} 
of the spin relaxation time in (110) quantum wells,  
while the modulation doping in barriers has also been employed 
in measurements\cite{Adachi2001, Dohrmann2004, Muller2008} and 
calculations.\cite{Zhou2009, Sherman2003}  

In this paper we adopt 
the delta-doping\cite{Ploog1987,Haug1987} in the well layer $(|z|<W/2)$. 
By the method of delta-doping, 
it is possible, in principle, to dope donor impurities in a particular atomic layer. 
We choose a doping symmetric with respect to the well center ($z=0$): 
a delta-doping on two atomic layers at $z= \pm z_{\rm d}$ $(z_{\rm d}<W/2)$.  
Such a symmetric doping  
keeps the impurity potential averaged over the plane symmetric. 
Note, however, that the impurity potential $V_{\rm imp}$ 
is not symmetric because of the random distribution of impurities in the plane. 

First we consider a delta-doping with $z_{\rm d}=0$ 
in which all impurities are on the center plane of the well (Fig.\ \ref{fig:delta_doping}(a)). 
In such a delta-doping  
the impurity potential $V_{\rm imp}$ is even in $z$. 
Terms with $\sigma_z$ in $H^{\rm so}_{\rm imp}$ and $H^{\rm so}_{\rm D}$ 
are also even in $z$,   
while terms with $\sigma_x$ or $\sigma_y$ 
in $H^{\rm so}_{\rm imp}$, $H^{\rm so}_{\rm well}$ and $H^{\rm so}_{\rm D}$ 
are odd in $z$, 
since $\hat p_z \rightarrow -\hat p_z$ when $z \rightarrow -z$. 

Such a symmetry with respect to $z=0$ leads to the absence of spin-flip scatterings 
with initial and final states in the same subband, 
which is valid in any orders of the impurity potential and the SOI. 
\footnote{
It has been shown in Eqs.\ (8) and (9) of our previous paper\cite{Akera2013} that  
the SOI induced by the impurity potential, $H_{\rm imp}^{\rm so}$, 
does not give any spin-flip matrix elements between states in the ground subband 
when all impurities are located on the center plane ($z = 0$) of a symmetric well. 
}  
Note that,  
since we have assumed that only the ground subband is occupied by electrons 
and consider only elastic processes, 
both initial and final states should be in the ground subband. 

\begin{figure}[ht]
\includegraphics[width=12cm, bb=0 0 842 595]{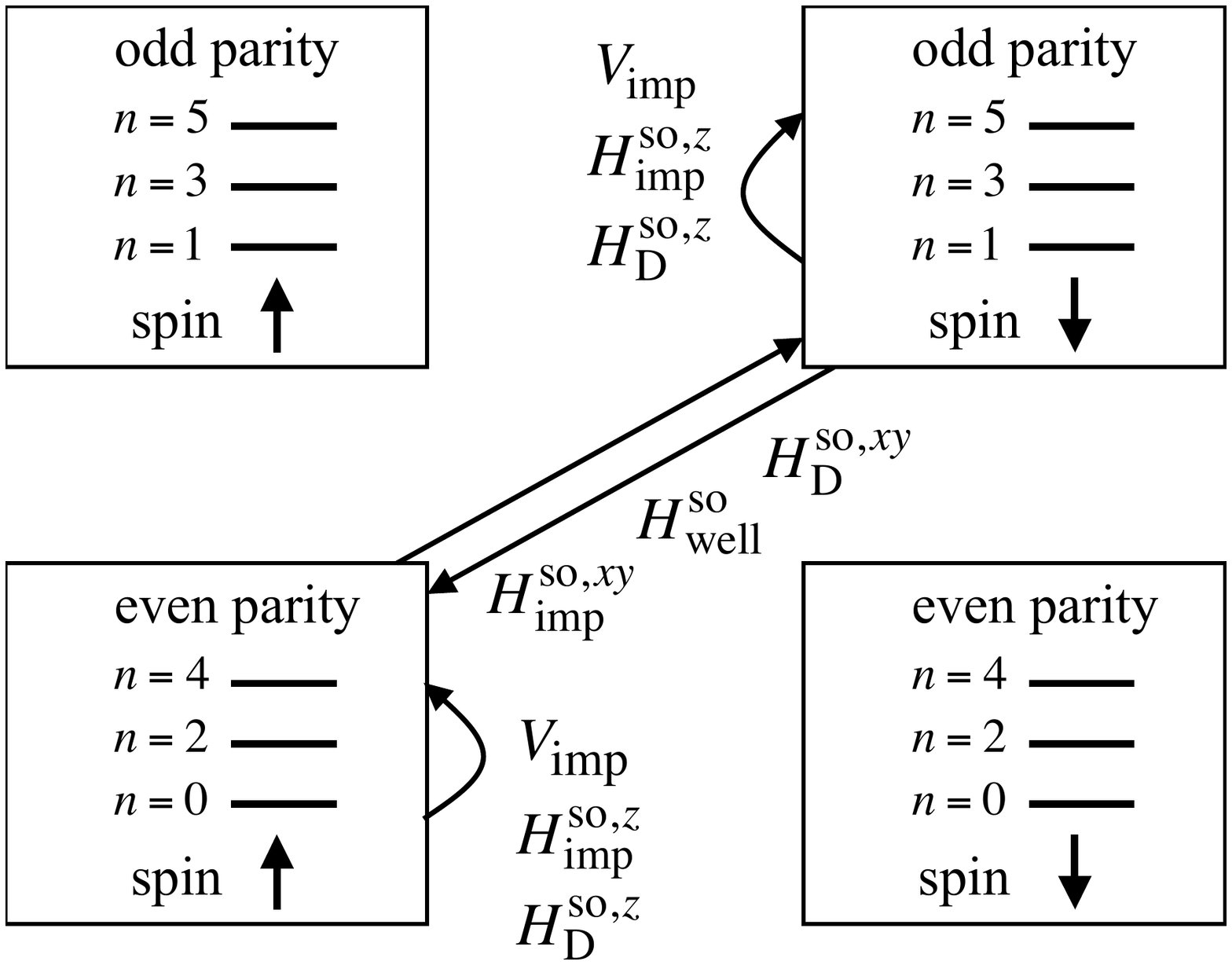}
\vskip -2.5cm
\caption{\label{fig:even_odd}
Each electron state has a parity in addition to a spin ($z$ component) 
since the well potential $V_{\rm well}$ is even in $z$. 
When $z_{\rm d}=0$ (Fig.\ \ref{fig:delta_doping}(a)) and 
the impurity potential $V_{\rm imp}$ is also even in $z$,   
some perturbation terms change the parity and the spin at the same time,  
while all others change neither the parity nor the spin. 
$H_{\rm imp}^{{\rm so}, xy}$ and $H_{\rm imp}^{{\rm so},z}$ 
($H_{\rm D}^{{\rm so}, xy}$ and $H_{\rm D}^{{\rm so},z}$)
denote terms with $\sigma_x$ or $\sigma_y$ and that with $\sigma_z$, respectively,  
of $H^{\rm so}_{\rm imp}$ ($H^{\rm so}_{\rm D}$).}
\end{figure}

The absence of such spin-flip scatterings is illustrated in Fig.\ \ref{fig:even_odd}. 
First note that each wave function associated with $z$ in a symmetric quantum well 
has a parity: 
even parity for $n=$even and odd parity for $n=$odd. 
Therefore each electron state is characterized by the parity and the spin $\sigma$. 
Terms in the perturbation $H_1$ with $z_{\rm d}=0$, 
which are odd in $z$ and include $\sigma_x$ or $\sigma_y$, 
change the parity and the spin at the same time, 
while all the others are even in $z$ with $\sigma_z$ and 
change neither the parity nor the spin. 
Since initial and final states of the considered spin-flip scattering processes 
have the same parity and the opposite spin, 
such processes do not occur by the perturbation $H_1$ with $z_{\rm d}=0$. 

\section{Spin-flip scatterings \\ in off-center delta-dopings}
\label{sec:off-center_delta-doping}

\subsection{Spin relaxation time in terms of \\ the spin-flip scattering rate}

Next we investigate the spin relaxation in the case of off-center delta-dopings 
with $z_{\rm d} \not= 0$ (Fig.\ \ref{fig:delta_doping}(b)).  
In this subsection we derive the formula of the spin relaxation time,  
which is given in terms of the spin-flip scattering rate of electrons.  

The spin polarization, 
or the $z$ component of the total spin angular momentum of the 2DES, 
is given by 
\begin{equation}
S_z= \sum_{n \Vec k}  \frac{\hbar}{2} 
\left( f_{n \Vec k \uparrow} - f_{n \Vec k \downarrow} \right),
\end{equation}
in terms of the occupation probability, $f_{n \Vec k \sigma}$, 
of a state with two-dimensional momentum $\Vec k $ and spin $\sigma$ in the $n$th subband. 
Our assumption that electrons occupy only the ground subband with $n=0$ is expressed by  
\begin{equation}
f_{n \Vec k \sigma}= 0 \ \ (n \ge 1) .
\end{equation}
We assume that $f_{0 \Vec k \sigma}$ is given by 
the Fermi distribution function with the spin-dependent chemical potential, 
$\mu_{\sigma}$: 
\begin{equation}
f_{0 \Vec k \sigma}= f_{\sigma} (\ve_{0k}) 
= \left[ \exp \left( \frac{\ve_{0k}-\mu_{\sigma}}{\kB T}\right) +1 \right]^{-1}. 
\end{equation}
Then the spin polarization becomes
\begin{equation}
S_z=  \frac{\hbar}{2} \int_{\ve_0}^{\infty} d\ve D
\left[ f_{\uparrow} (\ve) - f_{\downarrow} (\ve) \right] ,
\end{equation}
where $D$ 
is the constant density of states per spin of the ground subband 
and $\ve_0$ is the energy at the bottom of the ground subband at $\Vec k=0$.   

The spin polarization $S_z$ changes at each of spin-flip scatterings. 
With use of the transition rate, $W_{0 \Vec k' \bar \sigma \leftarrow  0 \Vec k \sigma}$, 
of a spin-flip scattering from $0 \Vec k \sigma$ to $0 \Vec k' \bar \sigma$  
($\bar \sigma$ is the spin opposite to $\sigma$),   
the time derivative of $S_z$ is  
\begin{equation}
\frac{d S_z}{dt}
= \sum_{\Vec k, \Vec k'} 
\hbar \left( 
- W_{0 \Vec k' \downarrow \leftarrow 0 \Vec k \uparrow} f_{0 \Vec k \uparrow} 
+ W_{0 \Vec k' \uparrow \leftarrow 0 \Vec k \downarrow} f_{0 \Vec k \downarrow}
\right)  .
\end{equation}
Here we define 
the total spin-flip scattering rate of an electron in a state $0 \Vec k \sigma$ by 
\begin{equation}
P^{\rm sf}_{0\Vec k \sigma} 
= \sum_{\Vec k'} W_{0 \Vec k' \bar \sigma \leftarrow  0 \Vec k \sigma} ,
\label{eq:Psf}
\end{equation}
and write the equation for the time derivative of $S_z$ as
\begin{equation}
\frac{d S_z}{dt}
= \sum_{\Vec k} \hbar \left( 
- P^{\rm sf}_{0\Vec k \uparrow} f_{0 \Vec k \uparrow}
+ P^{\rm sf}_{0\Vec k \downarrow} f_{0 \Vec k \downarrow}
\right) .
\label{eq:dSzdt2}
\end{equation}
Since $f_{0 \Vec k \sigma}=f_{\sigma} (\ve_{0k})$, 
it is convenient to separate the summation with respect to $\Vec k$ into 
the integration with respect to energy $\ve$ and 
the summation over the constant energy surface: 
\begin{equation}
\sum_{\Vec k} \cdots = 
\int_{\ve_0}^{\infty} d \ve \sum_{\Vec k} \delta(\ve-\ve_{0\Vec k}) \cdots . 
\label{eq:summation}
\end{equation}
In addition we introduce 
the average of $P^{\rm sf}_{0\Vec k \sigma}$ over the constant energy surface  
as
\begin{equation}
\bar P^{\rm sf}(\ve)
= \frac{1}{D}  \sum_{\Vec k} \delta(\ve - \ve_{0\Vec k}) P^{\rm sf}_{0\Vec k \sigma} .  
\label{eq:Psf_average}
\end{equation}
which is shown to be independent of spin. 
\footnote{
To show that $\bar P^{\rm sf}(\ve)$ in Eq.\ (\ref{eq:Psf_average}) 
is independent of $\sigma$,  
we write  
\[
\sum_{\Vec k} \delta(\ve - \ve_{0\Vec k}) P^{\rm sf}_{0\Vec k \sigma} 
=\sum_{\Vec k, \Vec k'} \delta(\ve - \ve_{0\Vec k})
W_{0 \Vec k' \bar \sigma \leftarrow  0 \Vec k \sigma} .
\]
and use the detailed balance: 
$W_{0 \Vec k' \bar \sigma \leftarrow  0 \Vec k \sigma}=
W_{0 \Vec k \sigma \leftarrow  0 \Vec k' \bar \sigma}$
and the energy conservation: $\ve_{0\Vec k}=\ve_{0\Vec k'}$. 
}
Equation (\ref{eq:dSzdt2}), with Eqs.\ (\ref{eq:summation}) and (\ref{eq:Psf_average}), reduces to 
\begin{equation}
\frac{d S_z}{dt}
= (-\hbar) \int_{\ve_0}^{\infty} d \ve D \bar P^{\rm sf}(\ve) 
 \left[ f_{\uparrow} (\ve) - f_{\downarrow} (\ve) \right] .
\label{eq:dSzdt3}
\end{equation}

Here we assume a degenerate 2DES 
satisfying $\kB T \ll \ve_{\rm F}-\ve_0$ 
($\ve_{\rm F}$: the Fermi energy) 
and a small spin polarization 
satisfying $|\mu_{\uparrow} - \mu_{\downarrow} | \ll \ve_{\rm F}-\ve_0$.   
Then $ f_{\uparrow} (\ve) - f_{\downarrow} (\ve)$ is negligibly small  
except in the close vicinity of $\ve_{\rm F}$, 
and Eq.\ (\ref{eq:dSzdt3}) becomes 
\begin{equation}
\frac{d S_z}{dt}
= (-\hbar) \bar P^{\rm sf}(\ve_{\rm F}) \int_{\ve_0}^{\infty} d \ve D  
 \left[ f_{\uparrow} (\ve) - f_{\downarrow} (\ve) \right] 
= - \frac{1}{\tau_{\rm s}} S_z ,
\end{equation}
with
\begin{equation}
\frac{1}{\tau_{\rm s}} = 2 \bar P^{\rm sf}(\ve_{\rm F})  .
\label{eq:spin_relaxation_time}
\end{equation}
Here $\tau_{\rm s}$ is the spin relaxation time.  
We have shown here that $1/\tau_{\rm s}$ is equal to twice  
the spin-flip scattering rate averaged over 
the Fermi surface of the 2DES.
\footnote{
The corresponding formula of the spin relaxation time in a bulk III-V semiconductor 
has appeared in 
Eq.\ (4) of Ref.\ \onlinecite{Chazalviel1975}.  
}

\subsection{Intrasubband and intersubband processes \\ giving spin-flip scatterings}

The transition rate appearing 
in the formula of the spin-flip scattering rate, Eq.\ (\ref{eq:Psf}), 
is given by 
\begin{equation}
W_{0 \Vec k' \bar \sigma \leftarrow  0 \Vec k \sigma} = 
\frac{2 \pi}{\hbar} 
\left| 
\left< 0 \Vec k' \bar \sigma \left| T \right| 0 \Vec k \sigma \right> 
\right|^2 
\delta(\ve_{0k'} - \ve_{0k} ) .
\end{equation}
Here $\left< 0 \Vec k' \bar \sigma \left| T \right| 0 \Vec k \sigma \right>$ 
is the transition matrix element.  

\begin{figure}[ht]
\includegraphics[width=9cm, bb=0 0 842 595]{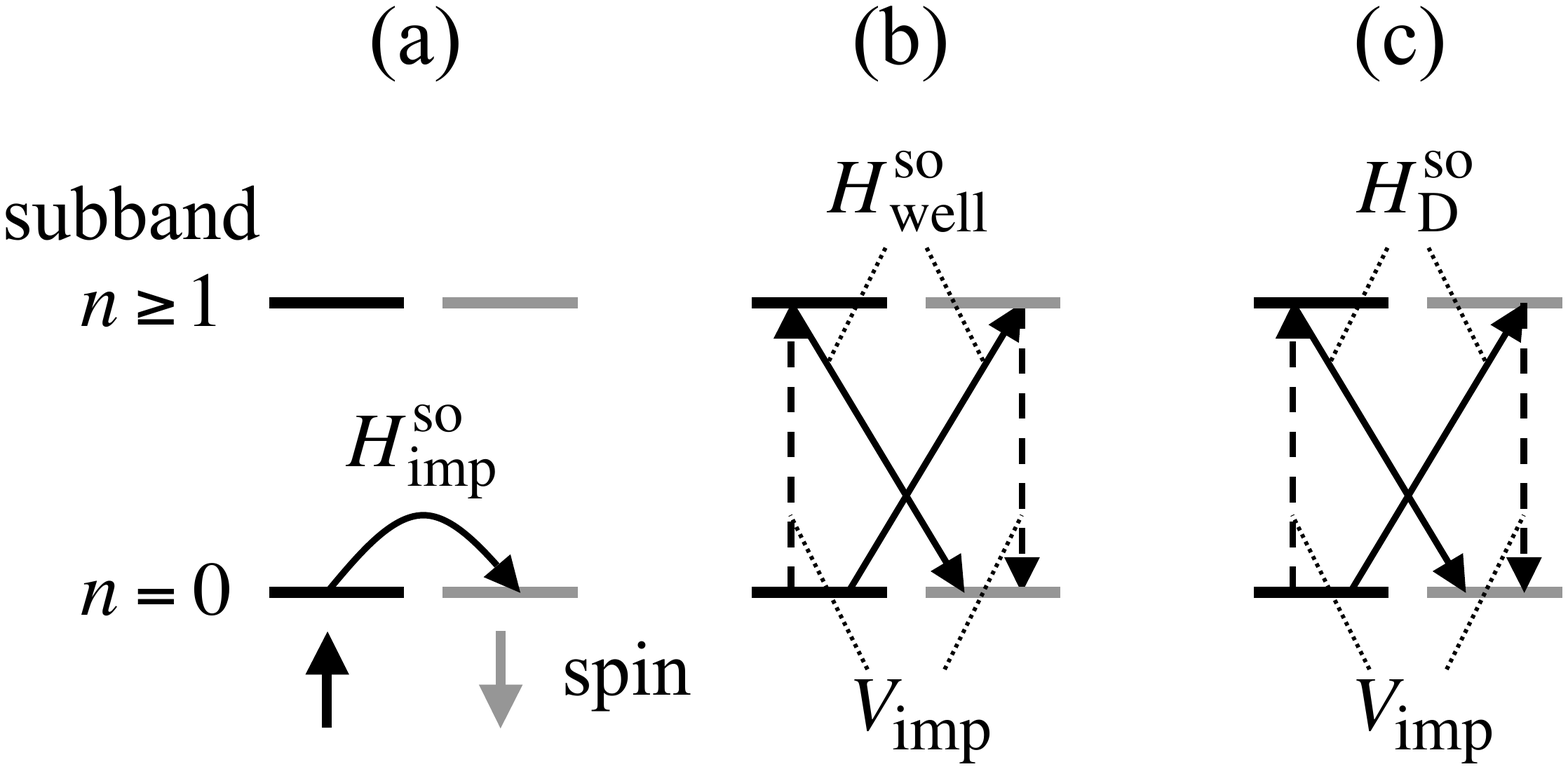}
\vskip -2.8cm
\caption{\label{fig:process}
(a) Intrasubband process 
by impurity-potential-induced SOI, $H^{\rm so}_{\rm imp}$. 
(b) Intersubband processes by well-potential-induced SOI, $H^{\rm so}_{\rm well}$, 
combined with impurity potential, $V_{\rm imp}$. 
(c) Intersubband processes by the Dresselhaus SOI, $H^{\rm so}_{\rm D}$,  
combined with $V_{\rm imp}$.
In (b) and (c), the summation is taken over excited subbands 
with odd parity $(n={\rm odd})$. 
}
\end{figure}

In deriving the transition matrix element, we take into account 
both intrasubband and intersubband processes 
with a spin flip 
by one of the SOIs, 
$H^{\rm so}_{\rm imp}$, $H^{\rm so}_{\rm well}$ and $H^{\rm so}_{\rm D}$. 
We retain terms of the transition matrix element 
in the lowest order both 
in the spin-orbit coupling strength represented by $\eta$ and $\gamma$ 
and in the impurity potential, $V_{\rm imp}$. 
A spin flip occurs due to the SOI  
and therefore requires at least the first order in $\eta$ or in $\gamma$. 
From the argument in Sec.\ \ref{sec:center_delta-doping}, 
in order to have a spin-flip scattering process 
with both initial and final states in the ground subband,  
we need to break the symmetry with respect to $z=0$ 
by introducing $V_{\rm imp}$ with $z_{\rm d} \not= 0$. 
Therefore the first order of $V_{\rm imp}$ is at least required. 
All of the processes, 
which are of the first order in the SOI and of the first order in $V_{\rm imp}$,  
are represented in Fig.\ \ref{fig:process}. 
The intrasubband process in Fig.\ \ref{fig:process}(a) 
is due to $H^{\rm so}_{\rm imp}$.  
The intersubband processes in Fig.\ \ref{fig:process}(b) and (c) are 
virtual processes through one of excited subbands 
caused by $H^{\rm so}_{\rm well}$ and $H^{\rm so}_{\rm D}$, respectively, 
combined with $V_{\rm imp}$. 

The transition matrix element 
for a spin-flip scattering from $0 \Vec k \sigma$ to $0 \Vec k' \bar \sigma$ 
consists of three terms, each corresponding to each process in Fig.\ \ref{fig:process}:
\begin{equation}
\left< 0 \Vec k' \bar \sigma \left| T \right| 0 \Vec k \sigma \right>
=T_{\rm intra}^{\Vec k' \bar \sigma \Vec k \sigma}
+ T_{\rm inter, well}^{\Vec k' \bar \sigma \Vec k \sigma}  
+ T_{\rm inter, D}^{\Vec k' \bar \sigma \Vec k \sigma}   , 
\end{equation}
where 
\begin{equation}
T_{\rm intra}^{\Vec k' \bar \sigma \Vec k \sigma}
= \left< 0 \Vec k' \bar \sigma \left| H^{\rm so}_{\rm imp} \right| 0 \Vec k \sigma \right> ,
\end{equation} 
\begin{equation}
\begin{split}
T_{\rm inter, well}^{\Vec k' \bar \sigma \Vec k \sigma} 
&= \sum_{n =1,3,\cdots} 
\frac{\left< 0 \Vec k' \bar \sigma \left| H^{\rm so}_{\rm well} \right| 
n \Vec k' \sigma \right>
\left< n \Vec k' \sigma
\left| V_{\rm imp} \right| 0 \Vec k \sigma \right>}
{\ve_0 - \ve_n} \\
&+ \sum_{n =1,3,\cdots} 
\frac{\left< 0 \Vec k' \bar \sigma \left| V_{\rm imp} \right| 
n \Vec k \bar \sigma \right>
\left< n \Vec k \bar \sigma 
 \left| H^{\rm so}_{\rm well} \right| 0 \Vec k \sigma \right>}
{\ve_0 - \ve_n} ,
\end{split}
\end{equation}
and $T_{\rm inter, D}^{\Vec k' \bar \sigma \Vec k \sigma}$ 
is obtained by replacing 
$H^{\rm so}_{\rm well}$ in $T_{\rm inter, well}^{\Vec k' \bar \sigma \Vec k \sigma}$ 
with $H^{\rm so}_{\rm D}$. 
Here the summation is taken over positive odd numbers and  
$\ve_n$ is the energy at the bottom of the $n$th subband. 

We assume that the impurity potential is the sum of contributions from each impurity:  
\begin{equation}
V_{\rm imp}(\Vec r)= \sum_j v(\Vec r - \Vec r_j) , 
\end{equation}
where $j$ labels each impurity, $\Vec r_j=(x_j,y_j,z_j)$ is the position of the $j$th impurity, 
and $v(\Vec r)$ is the potential created by an impurity 
when it is located at $\Vec r=0$. 
Then the intrasubband contribution becomes 
\begin{equation}
T_{\rm intra}^{\Vec k' \bar \sigma \Vec k \sigma} 
= \frac{\eta}{2S} K_{1\sigma}  \sum_j e^{-i \Vec q \cdot \Vec \rho_j}
\left< 0 \left| \left(\nabla_z \tilde v(q,z-z_j) \right) \right| 0 \right>
, 
\end{equation} 
where $\Vec \rho_j=(x_j,y_j)$, 
$\Vec q=(q_x, q_y)=\Vec k' - \Vec k$, 
$S$ is the area of the 2DES, and
\begin{equation}
K_{1\sigma}=  (k_y + k'_y) - i s_{\sigma} (k_x + k'_x), 
\end{equation}
with $s_{\sigma}=1$ ($\sigma =\uparrow$) and 
$s_{\sigma}=-1$ ($\sigma = \downarrow $).  
$\tilde v(q,z)$ with $q=(q_x^2 +q_y^2)^{1/2}$ 
is the two-dimensional Fourier transform of $v(\Vec r)$: 
\begin{equation}
\tilde v(q,z) = \int_{-\infty}^{\infty} dx \int_{-\infty}^{\infty} dy \ 
e^{-i \Vec q \cdot \Vec \rho}\ v(\Vec r) ,
\end{equation}
with  $\Vec \rho=(x,y)$. 
Since $v(\Vec r)$ depends on $\Vec \rho$ only through $| \Vec \rho |$, 
its two-dimensional Fourier transform has no dependence on the direction of $\Vec q$, 
and is real.  
The intersubband contributions are
\begin{equation}
\begin{split}
T_{\rm inter, well}^{\Vec k' \bar \sigma \Vec k \sigma} 
= &\frac{\eta}{S} K_{1\sigma} b_{\rm off} \sum_j e^{-i \Vec q \cdot \Vec \rho_j}  \\
 \sum_{n =1,3,\cdots}   &
\frac{\left< 0 \left| \left(\nabla_z V_{\rm well} \right) \right| n \right>
\left< n \left| \tilde v(q,z-z_j)  \right| 0 \right>}
{\ve_0 - \ve_n}
,
\end{split}
\end{equation} 
and
\begin{equation}
\begin{split}
T_{\rm inter, D}^{\Vec k' \bar \sigma \Vec k \sigma}
= \frac{\gamma}{S}& i K_{2\sigma} \sum_j e^{-i \Vec q \cdot \Vec \rho_j}  \\
 \sum_{n =1,3,\cdots}   &
\frac{\left< 0 \left| \nabla_z \right| n \right>
\left< n \left| \tilde v(q,z-z_j)  \right| 0 \right>}
{\ve_0 - \ve_n}
 ,
\end{split}
\label{eq:inter_D}
\end{equation} 
where 
\begin{equation}
K_{2\sigma}= 
\frac{k_x^2 - (k'_x)^2}{2} + k_y^2 - (k'_y)^2 - 2i s_{\sigma}  (k_x k_y - k'_x k'_y) .
\end{equation}

We first consider the case of $b_{\rm off}=1$, where 
$T_{\rm intra}^{\Vec k' \bar \sigma \Vec k \sigma}$ and 
$T_{\rm inter, well}^{\Vec k' \bar \sigma \Vec k \sigma}$ can be 
joined into
\begin{equation}
\begin{split}
T_{\rm intra}^{\Vec k' \bar \sigma \Vec k \sigma} 
&+T_{\rm inter, well}^{\Vec k' \bar \sigma \Vec k \sigma}
=  \frac{\eta}{2S}   K_{1\sigma}  \sum_j e^{-i \Vec q \cdot \Vec \rho_j}\\
 &
\left< \psi_0(z,z_j) \left| \left\{ \nabla_z \!
\left[  V_{\rm well} + \tilde v(q,z-z_j) \right] 
\right\} \right| \psi_0(z,z_j) \right>
. 
\end{split}
\label{eq:intra+inter_well}
\end{equation} 
Here
$\psi_0(z,z_j)$ is the ground-state wave function of a fictitious Hamiltonian, 
which includes a fictitious potential from a single impurity at $z_j$, $\tilde v(q,z-z_j)$: 
\begin{equation}
\left[ \frac{ \hat p_z^2}{2m} + V_{\rm well}(z) + \tilde v(q,z-z_j) \right] \psi_0(z,z_j) 
=\tilde \ve_0(z_j) \psi_0(z,z_j) , 
\end{equation}
where $\tilde \ve_0(z_j)$ is the corresponding energy eigenvalue.  
Note that the right hand side of Eq.\ (\ref{eq:intra+inter_well}) is to be evaluated 
in the first order of $\tilde v(q,z-z_j)$.  
We can show that each term of the right hand side of Eq.\ (\ref{eq:intra+inter_well}) is zero,  
since the average of the force induced by any potential $V(z)$ is zero 
when the average is taken with respect to the wave function $\psi(z)$ 
for each bound eigenstate 
of the Hamiltonian, $\hat p_z^2/2m + V(z)$, that is  
\begin{equation}
\left< \psi \left| \left(\nabla_z V \right) \right| \psi \right>=0 . 
\label{eq:zero_averaged_force}
\end{equation} 

The vanishing of $T_{\rm intra}^{\Vec k' \bar \sigma \Vec k \sigma}
+T_{\rm inter, well}^{\Vec k' \bar \sigma \Vec k \sigma}$ at $b_{\rm off}=1$ 
leads to its simplified formula at nonzero $b_{\rm off}$: 
\begin{equation}
T_{\rm intra}^{\Vec k' \bar \sigma \Vec k \sigma} 
+T_{\rm inter, well}^{\Vec k' \bar \sigma \Vec k \sigma}
=  (1-b_{\rm off}) T_{\rm intra}^{\Vec k' \bar \sigma \Vec k \sigma} .
\label{eq:intra+inter_well_2}
\end{equation} 
This equation shows that 
the intrasubband and intersubband terms interfere destructively 
when $b_{\rm off}>0$ and 
the interference becomes completely destructive at $b_{\rm off}=1$.
The formula of $b_{\rm off}$ given in Appendix shows that 
$b_{\rm off}$ can take a value close to unity in some type-II quantum wells.  

The same equality as Eq.\ (\ref{eq:zero_averaged_force}) has  been employed 
by Ando\cite{Darr1976,Ando1982} to show that 
the spin splitting, linear in the in-plane momentum of the 2DES, due to the SOI is absent 
when the SOI is proportional to $\nabla_z V$ 
where $V(z)$ is the confining potential of the 2DES, 
even if $V(z)$ has no inversion symmetry. 
Later the $\Vec k \cdot \Vec p$ theory developed for heterostructures
\cite{Lassnig1985, Silva1997, Pfeffer1999} 
has shown that the spin splitting is present when 
differences in   
the band gap and the spin-orbit splitting 
between the well and barrier layers are considered. 
This is because $b_{\rm off}\not= 1$, in general, and therefore  
the combined SOI due to the band offset and the electrostatic potential 
is not proportional to $\nabla_z V$ (see Appendix).

\subsection{Spin-flip scattering rate averaged \\ over impurity in-plane positions}

In calculating the spin-flip scattering rate 
averaged with respect to the direction of $\Vec k$, 
$\bar P^{\rm sf}(\ve)$, 
defined by Eq.\ (\ref{eq:Psf_average}) with Eq.\ (\ref{eq:Psf}),  
we perform another averaging of $\bar P^{\rm sf}(\ve)$ 
over various impurity configurations with the same doping position $z_{\rm d}$. 
This is performed by taking the average of 
$|\left< 0 \Vec k' \bar \sigma \left| T \right| 0 \Vec k \sigma \right>|^2$ 
over uncorrelated in-plane positions of impurities: 
\begin{equation}
\left( \prod_j \frac{1}{S} \int_S dx_j dy_j \right) 
|\left< 0 \Vec k' \bar \sigma \left| T \right| 0 \Vec k \sigma \right>|^2 .
\end{equation}

Then the spin-flip scattering rate $\bar P^{\rm sf}(\ve)$ is obtained to be  
\begin{equation}
\begin{split}
\bar P^{\rm sf}(\ve) = P_0 
\int_0^{2\pi} & d\theta  
\ [\  
(kW)^2 (1+ \cos \theta) t_{\rm pot}(q, z_{\rm d})^2 \\
&+ a_{\rm D} (kW)^4  (1- \cos 2\theta) t_{\rm inter}^{\rm D} (q, z_{\rm d})^2
\ ] , 
\end{split}
\label{eq:calculatedPsf}
\end{equation}
where $\ve=\hbar^2 k^2/2m +\ve_0$, $q=k\sqrt{2(1-\cos \theta)}$, 
$\theta$ is the angle of $\Vec k'$ with respect to $\Vec k$, and 
\begin{equation}
P_0 = \frac{\pi}{2\hbar} \frac{(e^2/\epsilon)^2 n_{\rm imp}}{\ve_0}
\left( \frac{\eta}{W^2} \right)^2 ,
\label{eq:P0}
\end{equation}
with $\epsilon$ the static dielectric constant of the semiconductor and 
$n_{\rm imp}$ the area density of impurities. 
The dimensionless parameter, $a_{\rm D}$, is defined by 
\begin{equation}
a_{\rm D}= \frac{17}{32} \left( \frac{\gamma}{\eta W \ve_0} \right)^2 ,
\label{eq:aD}
\end{equation}
with the ratio between $\gamma$, the coupling constant of the Dresselhaus SOI,  
and $\eta$, that of the potential-induced SOIs. 
The dimensionless quantity, $t_{\rm pot}(q, z_{\rm d})$, 
comes from terms of the transition matrix element caused by the potential-induced SOIs 
and is given, using $T_{\rm intra}^{\Vec k' \bar \sigma \Vec k \sigma}
+T_{\rm inter, well}^{\Vec k' \bar \sigma \Vec k \sigma}$ 
in Eq.\ (\ref{eq:intra+inter_well_2}), by
\begin{equation}
t_{\rm pot} (q, z_{\rm d})= 
\frac{\epsilon}{2e^2} \left< 0 \left| \left(\nabla_z \tilde v(q, z-z_{\rm d}) \right)  \right| 0 \right> 
(1-b_{\rm off}) .  
\end{equation}
On the other hand, 
$t_{\rm inter}^{\rm D}(q, z_{\rm d})$ 
is the contribution of the intersubband process due to $H^{\rm so}_{\rm D}$ 
and is given,  
from $T_{\rm inter, D}^{\Vec k' \bar \sigma \Vec k \sigma}$ 
in Eq.\ (\ref{eq:inter_D}), by
\begin{equation}
t_{\rm inter}^{\rm D} (q, z_{\rm d})=  
\ve_0\frac{\epsilon}{e^2} \sum_{n =1,3,\cdots} 
\frac{\left< 0 \left| \nabla_z \right| n\right>  \left< n \left| \tilde v(q, z-z_{\rm d})  \right| 0 \right>}
{\ve_0 -\ve_n} .  
\label{eq:t_D}
\end{equation}

We simplify the calculation of 
$t_{\rm pot}(q, z_{\rm d})$ and $t_{\rm inter}^{\rm D}(q, z_{\rm d})$ 
by taking the limit
\footnote{Although $b_{\rm off}$ defined by Eq.\ (\ref{eq:cso}) depends on $V_0=\Delta E_{\rm c}$, 
$b_{\rm off}$ is kept constant when the limit of $V_0 \rightarrow \infty$ is taken. 
This means that $\Delta E_{\rm v}$ and $\Delta E_{\rm so}$ increase at the same rate as $V_0$.} of $V_0 \rightarrow \infty$,  
which gives 
\begin{equation}
\ve_n = \frac{\hbar^2}{2m} \left[ \frac{(n+1) \pi}{W} \right]^2 ,
\end{equation}
and
\begin{equation}
\left< 0 \left| \nabla_z \right| n\right> = - \frac{4}{W} \frac{n+1}{(n+1)^2-1}
\ \ (n={\rm odd}) .  
\end{equation}
The potential of each donor impurity, $v(\Vec r)$, 
is modeled by a screened Coulomb potential:  
\begin{equation}
v(\Vec r)= - \frac{e^2}{\epsilon r} \exp(- k_{\rm s} r) ,
\label{eq:screened_Coulomb}
\end{equation}
where $r=|\Vec r|$ and $k_{\rm s}$ is the inverse of the screening length. 
Its two-dimensional Fourier transform is  
\begin{equation}
\tilde v(q, z)= - \frac{2\pi e^2}{\epsilon Q} \exp(- Q |z| )  ,  
\label{eq:v_q}
\end{equation}
with $Q=(q^2 +k_{\rm s}^2)^{1/2}$.

\subsection{Calculated spin-flip scattering rate \\ as a function of the impurity distribution}

We present the spin-flip scattering rate calculated 
for a quantum well made of GaAs and Al$_{0.4}$Ga$_{0.6}$As 
with the width $W=75\ {\rm \AA}$,  
as in the sample employed 
in the measurement by Ohno {\it et al.}\cite{Ohno1999} 
We use the following values of parameters for GaAs: 
$\gamma=27.5\ {\rm eV \AA}^3$ (Table III of Ref.\ \onlinecite{Knap1996}) and 
$m=0.067m_0$ with $m_0$ the electron rest mass. 
We obtain the value of $b_{\rm off}$ in Eq.\ (\ref{eq:cso})
and that of $\eta$ in Eq.\ (\ref{eq:eta})
to be $b_{\rm off}=-0.82$ for GaAs/Al$_{0.4}$Ga$_{0.6}$As quantum well 
and $\eta=5.28\ {\rm \AA}^2$ for GaAs
by using the band parameters\cite{Vurgaftman2001} of 
GaAs, AlAs, AlGaAs, and GaAs/AlAs. 
By substituting the values of $\gamma$, $m$, and $\eta$  
with $W=75\ {\rm \AA}$ into Eq.\ (\ref{eq:aD}), 
we obtain $a_{\rm D}=0.26$.

\begin{figure}[ht]
\includegraphics[width=10.5cm, bb=0 0 595 842]{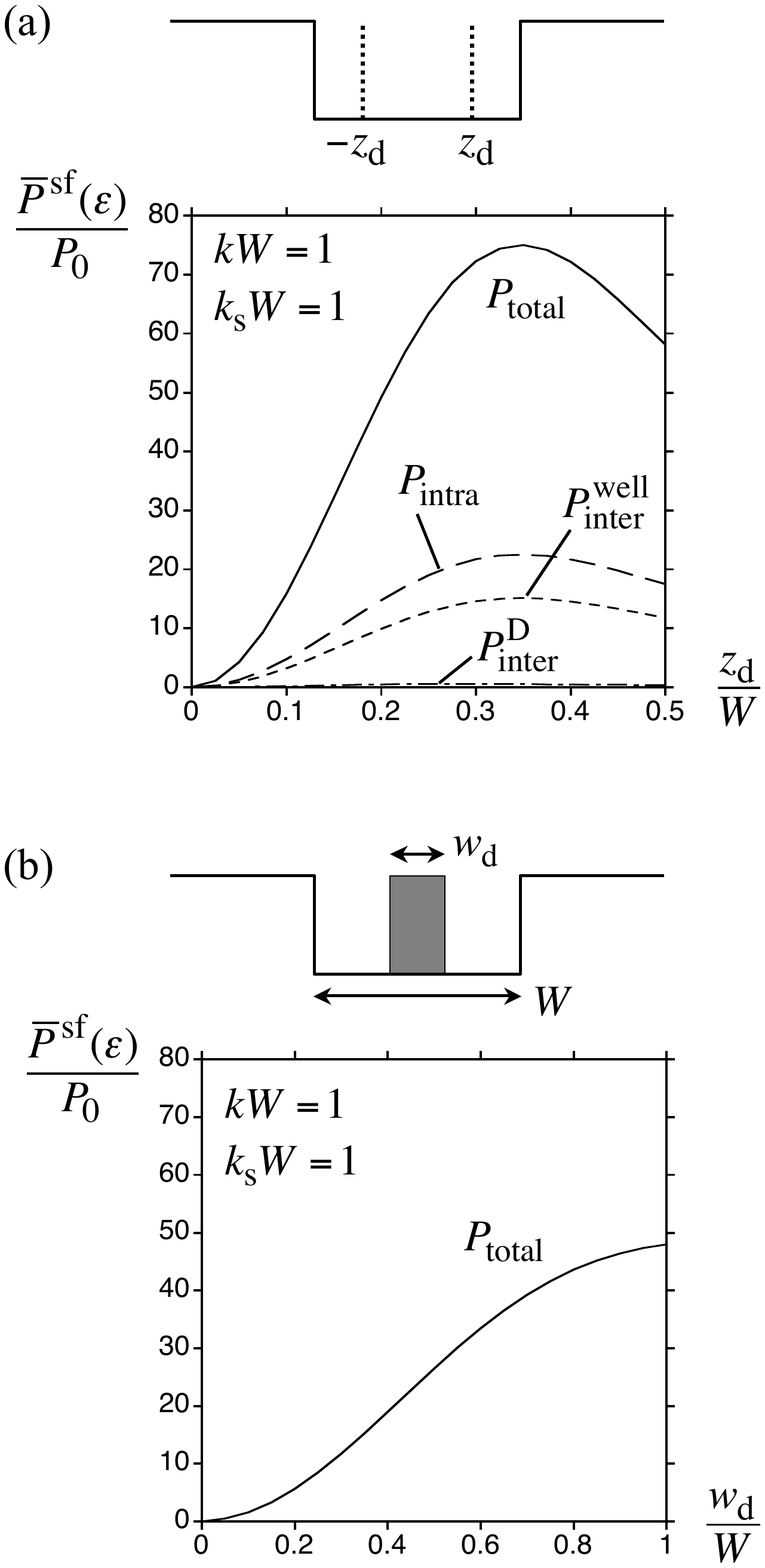}
\vskip -0.5cm
\caption{\label{fig:zd+wd_dep}
Spin-flip scattering rate, $\bar P^{\rm sf}(\ve)$, 
divided by $P_0$ (Eq.\ (\ref{eq:P0})). 
$k=|\Vec k|$ with $\Vec k =(k_x,k_y)$ and 
$k_{\rm s}$ is the inverse of the screening length. 
(a) Dependence on the position of delta-doping, $z_{\rm d}$. 
Contributions from each spin-flip scattering process in Fig.\ \ref{fig:process} 
are also shown.
(b) Dependence on the width of the doped layer, $w_{\rm d}$,  
with all of the three processes in Fig.\ \ref{fig:process} considered. 
}
\end{figure}

Figure \ref{fig:zd+wd_dep}(a) presents 
the calculated $z_{\rm d}$ dependence of $\bar P^{\rm sf}(\ve)$ 
for $k W=1$ and $k_{\rm s} W=1$, where 
$k_{\rm F} W=1$ with $W=75\ {\rm \AA}$ corresponds to 
the electron density of $2.8 \times 10^{11} {\rm cm}^{-2}$. 
Each of the curves labeled $P_{\rm intra}$, $P_{\rm inter}^{\rm well}$, and $P_{\rm inter}^{\rm D}$  
shows the value of $\bar P^{\rm sf}(\ve)$ 
when one of the processes, (a), (b), and (c), respectively in Fig.\ \ref{fig:process}, is considered. 
$P_{\rm intra}$, the intrasubband contribution,  
and $P_{\rm inter}^{\rm well}=b_{\rm off}^2 P_{\rm intra}$, the intersubband contribution due to $H^{\rm so}_{\rm well}$,   
are comparable in magnitude since $b_{\rm off}=-0.82$.  
$P_{\rm inter}^{\rm D}$, the intersubband contribution due to $H^{\rm so}_{\rm D}$,  
is about 0.03 of $P_{\rm inter}^{\rm well}$ in magnitude at the maximum. 
The value of $\bar P^{\rm sf}(\ve)$ when all of the three processes are considered   
is also plotted in Fig.\ \ref{fig:zd+wd_dep}(a) as a curve labeled $P_{\rm total}$. 
Since $b_{\rm off}$ is negative, the intrasubband and intersubband terms 
of $t_{\rm pot}(q, z_{\rm d})$
interfere constructively 
and therefore 
$P_{\rm total}$ in Fig.\ \ref{fig:zd+wd_dep}(a) is nearly four times larger 
than each of $P_{\rm intra}$ and $P_{\rm inter}^{\rm well}$.  
$\bar P^{\rm sf}(\ve)$ as a function of $z_{\rm d}$ increases in the vicinity of the well center, 
while it decreases near the well boundary 
because the expectation value and the matrix element 
of the screened Coulomb potential, Eq.\ (\ref{eq:screened_Coulomb}), 
are reduced in magnitude. 

Next we consider impurity distributions with nonzero widths:   
impurities are distributed uniformly within a layer in $-w_{\rm d}/2 < z < w_{\rm d}/2$. 
We change the width of the doped layer, $w_{\rm d}$, 
with the total number of impurities kept constant. 
Figure \ref{fig:zd+wd_dep}(b) shows a calculated result of 
$\bar P^{\rm sf}(\ve)$ as a function of $w_{\rm d}$.  
$\bar P^{\rm sf}(\ve)$ remains small for small values of $w_{\rm d}$ and
increases monotonically with $w_{\rm d}$. 
Suppose that a possible diffusion of impurities from the delta-doped layer 
gives an impurity distribution over three (110) atomic layers. 
Then $w_{\rm d}$ is twice the atomic layer distance: 
$w_{\rm d} =a/\sqrt 2 =4.0\ {\rm \AA}$ for GaAs 
with $a=5.65\ {\rm \AA}$. 
$\bar P^{\rm sf}(\ve)$ for this value of $w_{\rm d}$ is found to be 
two orders of magnitude smaller than that for $w_{\rm d}=W$ 
(uniform distribution in the full width of the well) 
when $W=75\ {\rm \AA}$. 

\begin{figure}[h]
\includegraphics[width=8.5cm, bb=0 0 595 842]{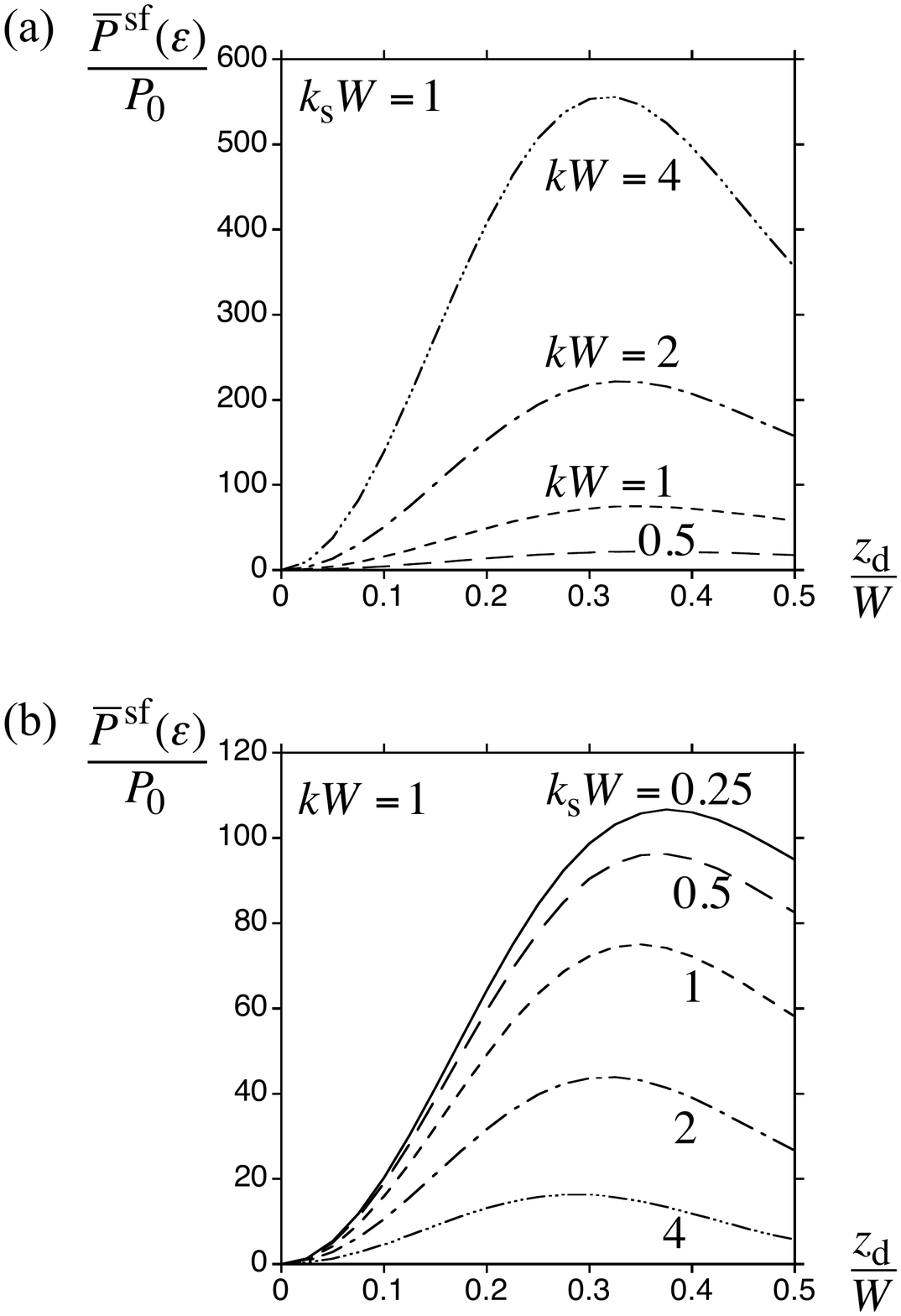}
\vskip -1.3cm
\caption{\label{fig:k+ks_dep}
Spin-flip scattering rate, $\bar P^{\rm sf}(\ve)$, 
divided by $P_0$ (Eq.\ (\ref{eq:P0})), 
as a function of the position of delta-doping, $z_{\rm d}$. 
(a) Dependence on $k=|\Vec k|$. 
(b) Dependence on $k_{\rm s}$, the inverse of the screening length. 
}
\end{figure}

Figure \ref{fig:k+ks_dep}(a) demonstrates  
the dependence of $\bar P^{\rm sf}(\ve)$ on the electron momentum, $k$:  
$\bar P^{\rm sf}(\ve)$ increases as $kW$ becomes larger. 
The origin of this increase is 
the factor $(kW)^2$ in front of $t_{\rm pot}(q, z_{\rm d})^2$ in Eq.\ (\ref{eq:calculatedPsf}), 
which is partly suppressed by the $k$ dependence of $\tilde v(q, z)$ in Eq.\ (\ref{eq:v_q}) 
through $q^2=2k^2(1-\cos \theta)$. 

Figure \ref{fig:k+ks_dep}(b) shows 
the dependence of $\bar P^{\rm sf}(\ve)$ 
on the inverse of the screening length, $k_{\rm s}$: 
$\bar P^{\rm sf}(\ve)$ decreases with $k_{\rm s}$. 
This comes from the $k_{\rm s}$ dependence of $\tilde v(q, z)$ in Eq.\ (\ref{eq:v_q})
through $Q=(q^2 +k_{\rm s}^2)^{1/2}$. 
$\bar P^{\rm sf}(\ve)$ approaches a constant value as $k_{\rm s} \rightarrow 0$, 
since $Q\rightarrow q$ then.

\section{Conclusions}
\label{sec:conclusions}

We have investigated theoretically the dependence of the spin relaxation rate 
on the impurity distribution in a zinc-blende (110) symmetric quantum well 
for the spin orientation perpendicular to the well, 
by calculating the spin-flip scattering rate.  
First we have considered a delta-doping on the center plane of the well at $z=0$ 
and shown that the symmetry with respect to $z=0$
of the impurity potential and the well potential 
leads to the vanishing of all spin-flip scattering processes 
when only the ground subband is occupied by electrons. 

Next we have considered the presence of impurities 
in positions deviated from the well center.  
We have found that the spin-flip scattering rate remains small for 
narrow impurity distributions centered at $z=0$:  
the spin-flip scattering rate for the distribution width of 4\ \AA \ 
(twice the distance between adjacent GaAs (110) atomic layers) is 
estimated to be two orders of magnitude smaller than 
that for the uniform distribution over the well width of 75\ \AA. 

In the calculation we have taken into account all processes, which are 
in the first order of the SOI and, at the same time,  
in the first order of the impurity potential. 
We have found that the intersubband spin-flip scattering process due to 
the well-potential-induced SOI gives a contribution 
comparable to the intrasubband process. 
In type-II quantum wells, 
the interference between these two processes can be destructive, 
which may result in a strong suppression of the spin relaxation. 
In type-I quantum wells made of GaAs and AlGaAs, however, 
these two processes interfere constructively, 
giving an enhanced spin-flip scattering rate,  
while the third contribution from 
the intersubband process caused by the Dresselhaus SOI 
makes only a negligible contribution.


\appendix*
\section{}

Here we derive the formula of $b_{\rm off}$ appearing in Eq.\ (\ref{eq:SOI_well}), 
by following the $\Vec k \cdot \Vec p$ theory developed for 
heterostructures.\cite{Lassnig1985, Bernardes2007,Calsaverini2008}  
The potential acting on an electron is 
due to either the band offset or the electrostatic potential.   
The potential due to the band offset at the interface of heterostructures   
depends on the band which the electron occupies.  
Without specifying whether it is due to the band offset or the electrostatic potential 
until Eq.\ (\ref{eq:electrostatic}) below,  
we denote 
the potential acting on an electron in the conduction band by $V_{\rm c}(z)$, 
that in the heavy-hole plus light-hole bands by $V_{\rm v}(z)$,   
and that in the split-off band by $V_{\rm so}(z)$. 
The SOI for an electron in the conduction band 
is induced by position dependences of $V_{\rm v}(z)$ and $V_{\rm so}(z)$ 
through the mixing between the conduction and valence bands 
by the $\Vec k \cdot \Vec p$ term, 
and is given for an electron with momentum $(k_x, k_y)$ 
by\cite{Lassnig1985, Bernardes2007, Calsaverini2008}
\begin{equation}
H_{\rm so}= \frac{P^2}{3} 
\left[ 
\frac{\nabla_z V_{\rm v}}{E_{\rm g}^2}  
- \frac{\nabla_z V_{\rm so}}{(E_{\rm g}+ \Delta_{\rm so})^2}  
\right]
\left(
\sigma_x k_y - \sigma_y k_x
\right) ,
\label{eq:Hso}
\end{equation}
where $E_{\rm g}$ is the band gap and    
$\Delta_{\rm so}$ is the spin-orbit splitting. 
$P$ is the Kane matrix element\cite{Kane1957} given by 
\begin{equation}
P= - i \frac{\hbar}{m_0} \left< S|\hat p_x|X\right> .
\end{equation}
Here $m_0$ is the electron rest mass, 
while $|S \rangle$ and $|X\rangle$ are 
the s-type wave function at the conduction-band bottom and 
the p-type wave function at the valence-band top, respectively. 

First consider the case of the electrostatic potential. 
In this case $V_{\rm c}(z)$, $V_{\rm v}(z)$ and $V_{\rm so}(z)$ 
are all equal to the electrostatic potential energy $V_{\rm es}(z)$: 
\begin{equation}
V_{\rm c}(z)= V_{\rm v}(z) = V_{\rm so}(z) = V_{\rm es}(z) .
\label{eq:electrostatic}
\end{equation}
Then Eq.\ (\ref{eq:Hso}) becomes 
\begin{equation}
H_{\rm so}= \eta \ (\nabla_z V_{\rm es})
\left(
\sigma_x k_y - \sigma_y k_x
\right) ,
\end{equation}
with 
\begin{equation}
\eta = \frac{P^2}{3} 
\left[ 
\frac{1}{E_{\rm g}^2}  
- \frac{1}{(E_{\rm g}+ \Delta_{\rm so})^2}  
\right] ,
\label{eq:eta}
\end{equation}
which gives the formula of the effective coupling constant of the SOI 
appearing in Eq.\ (\ref{eq:SOI_imp}). 

Next we consider the case where the potentials are due to the band offset. 
In a quantum well with width $W$  
\begin{equation}
\begin{split}
V_{\rm c}(z)&=\Delta E_{\rm c} h(z)=V_{\rm well}(z), \\ 
V_{\rm v}(z)&=\Delta E_{\rm v} h(z), \\ 
V_{\rm so}(z)&=\Delta E_{\rm so} h(z), 
\end{split}
\end{equation}
with 
\begin{equation}
h(z)=
\left\{ \begin{array}{ll}
                           0  &(|z|<W/2) ,   \\ 
                           1  &(|z|>W/2) .  
           \end{array}   \right.   
\end{equation}
Here $\Delta E_{\rm c}(=V_0)$, $\Delta E_{\rm v}$, and $\Delta E_{\rm so}$ 
are band offsets of the corresponding bands,  
defined by the offset of the energy in the barrier layers relative to that in the well layer.  
Introducing $E_{\rm c}$ ($E_{\rm c}^{\rm b}$) the energy of the conduction-band bottom, 
$E_{\rm v}$ ($E_{\rm v}^{\rm b}$) that of the valance-band top, 
$E_{\rm so}$ ($E_{\rm so}^{\rm b}$) that of the split-off-band top 
in the well layer (the barrier layers),   
we have $E_{\rm g}= E_{\rm c}- E_{\rm v} $,   
$\Delta_{\rm so}= E_{\rm v}- E_{\rm so} $, 
$\Delta E_{\rm c} = E_{\rm c}^{\rm b}- E_{\rm c}$,  
$\Delta E_{\rm v} = E_{\rm v}^{\rm b}- E_{\rm v} $, and  
$\Delta E_{\rm so} = E_{\rm so}^{\rm b}- E_{\rm so}$. 
In this case Eq.\ (\ref{eq:Hso}) becomes
\begin{equation}
H_{\rm so}= \eta b_{\rm off} (\nabla_z V_{\rm c})
\left(
\sigma_x k_y - \sigma_y k_x
\right) ,
\end{equation}
with the formula of $b_{\rm off}$: 
\begin{equation}
 b_{\rm off} =  \frac{\Delta E_{\rm v} E_{\rm g}^{-2} -\Delta E_{\rm so} (E_{\rm g}+ \Delta_{\rm so})^{-2}}
 {\Delta E_{\rm c} \left[E_{\rm g}^{-2} - (E_{\rm g}+ \Delta_{\rm so})^{-2} \right]} . 
\label{eq:cso}
\end{equation}

\bibliography{Spin_Relaxation}

\end{document}